%% file: main.tex
\documentclass{aa}

\input{style/settings}

\usepackage{graphicx}
\usepackage{newtxtext}
\usepackage[smallerops,cmbraces]{newtxmath}
\usepackage[
  colorlinks=true,        
  linkcolor=xlinkcolor,   
  citecolor=xlinkcolor,   
  filecolor=xlinkcolor,   
  urlcolor=xlinkcolor
]{hyperref}

\input{style/packages}
\input{style/definitions}

\begin{document}
\definecolor{xlinkcolor}{cmyk}{1,1,0,0}

\title{Sensitivity of halo shape measurements}
\titlerunning{Shape Sensitivity}

\author{
    Moritz S.\ Fischer\inst{\ref{inst:usm},\ref{inst:origins},\ref{inst:uhh}}
    \and
    Lucas M.\ Valenzuela\inst{\ref{inst:usm}}
}
\authorrunning{M.\ S.\ Fischer \& L.\ M.\ Valenzuela}

\institute{
    Universitäts-Sternwarte, Fakultät für Physik, Ludwig-Maximilians-Universität München, Scheinerstr.\ 1, D-81679 München, Germany\label{inst:usm}\\
    \email{mfischer@usm.lmu.de}
    \and
    Excellence Cluster ORIGINS, Boltzmannstrasse 2, D-85748 Garching, Germany\label{inst:origins}
    \and
    Hamburger Sternwarte, Universität Hamburg, Gojenbergsweg 112, D-21029 Hamburg, Germany\label{inst:uhh}
}

\date{Received XX Month, 20XX / Accepted XX Month, 20XX}

\abstract{
Shape measurements of galaxies and galaxy clusters are widespread in the analysis of cosmological simulations. But the limitations of those measurements have been poorly investigated. In this paper, we explain why the quality of the shape measurement does not only depend on the numerical resolution, but also on the density gradient. In particular, this can limit the quality of measurements in the central regions of haloes. We propose a criterion to estimate the sensitivity of the measured shapes based on the density gradient of the halo and to apply it to cosmological simulations of collisionless and self-interacting dark matter. By this, we demonstrate where reliable measurements of the halo shape are possible and how cored density profiles limit their applicability.}


\maketitle
%

\section{Introduction} \label{sec:introduction}

The shapes of galaxies and more massive objects such as galaxy clusters have been measured in numerous publications.
In simulations, it is simple to measure the shape of various components such as stars, gas, or dark matter (DM).

The shape of astrophysical objects is of interest because the potential directly depends on it, influencing the orbits and movement of DM, stars, gas, and satellite galaxies around the host system.
Shapes can be affected by various physical mechanisms and thus measurements of the shape potentially allow us to learn about such processes.
For instance, \cite{Chua_2022} studied the impact of galactic feedback on the shape of DM haloes.
Shapes were also investigated in studies of DM, for example, self-interacting dark matter (SIDM) tends on average to make haloes more round \citep[e.g.][]{Peter_2013}.

Shapes have been measured in $N$-body simulations for a long time, starting with early DM-only simulations \citep[e.g.][]{Katz_1991} and the first larger cosmological simulations \citep[e.g.][]{Springel_2004, Allgood_2006, Bett_2007} studying the DM halo shapes and their relations to other properties, such as mass and spin.
Such studies found that DM haloes are generally prolate, that is elongated in shape.
Furthermore, it has been shown that more massive systems are on average more ellipsoidal \citep[e.g.][]{Jing_2002, Allgood_2006, Munoz-Cuartas_2011, Despali_2013, Despali_2014}.

More recent work has been done on using cosmological simulations to study the impact of baryons and the cosmic web environment on shapes \citep[e.g.][]{Abadi_2010, Kazantzidis_2010, Tissera_2010, Zemp_2012, Bryan_2013, Butsky_2016, Chua_2019, Cataldi_2021, Hellwing_2021}.
In particular, \citet{Bonamigo_2015} give a good overview about halo shapes in cosmological simulations, comparing the results of multiple studies.
Also, the stellar component and its shape in hydrodynamical simulations have been studied in recent years \citep[e.g.][]{Pulsoni_2020, Emami_2021b, Valenzuela_2022} and compared with the respective DM shapes and their alignments to each other \citep[e.g.][]{Tenneti_2014,Velliscig_2015}.

There exist a variety of methods to infer the shapes from $N$-body simulations.
A number of these methods have been studied and compared by \cite{Zemp_2011}, and additional methods have also been applied in other works \citep[e.g.][]{Warnick_2008, Emami_2021a}.
However, the limitations of the sensitivity of the shape measurements, and with their accuracy, have been poorly studied.
Typically, one only uses lower thresholds for the number of particles that should be included in the ellipsoidal shell used to determine the shape.
For example, \citet{Zemp_2011} recommended requiring at least a few thousand particles to lie within each shell.
Other authors have also set lower thresholds of 1000 particles per shell \citep[e.g.][]{Pulsoni_2020}.
In this paper, we point out that a large number of particles is not sufficient to measure accurate shapes, but that the density gradient matters as well.

In general, shape measurements roughly trace isodensity contour lines.
In the case of a constant density, isodensity contours would no longer exist and the shape would become undefined.
Therefore, we study in this work how crucial the steepness of the density gradient is for measuring shapes.
In Sect.~\ref{sec:shape_gradient} we explicitly demonstrate that the shape is becoming undefined and thus the relevance of the density gradient is inherent to the definition of halo shapes.

It is known that haloes can have density cores, that is, their central density is roughly constant.
Mechanisms that can reduce the central density to be constant include supernova feedback and DM self-interactions.
Given such a density core, state-of-the-art algorithms can nevertheless return a shape, though it is anything but reliable.
Here, the discretisation noise of the $N$-body simulations starts to play a crucial role, as we show in Sect.~\ref{sec:shape_crit}.
If it is larger than the density gradient, it may dominate the shape measurement and the obtained shape can be far off from the physical shape.
In this case, it would typically return shapes that are too elliptical.
To understand this quantitatively, we developed a criterion to measure how accurate the inferred shapes are.
Such a criterion is of interest for any studies that rely on the measurement of shapes, for example in the context of DM \citep[e.g.][]{Despali_2022, Shen_2022} or the impact of baryons \citep[e.g.][]{Springel_2004, Prada_2019}.

In this section, we have described the problem at hand. 
The following section (Sect.~\ref{sec:shapes}) deals with how we measured shapes and we introduced the criterion to estimate the sensitivity of the shape measurement.
In Sect.~\ref{sec:results}, we explain how we measured the shape of simulated haloes and studied the effect of density cores.
Subsequently, in Sect.~\ref{sec:discussion}, we discuss our findings.
Finally, we conclude in Sect.~\ref{sec:conclusion}.

\section{Shapes} \label{sec:shapes}

In this section, we describe how we measured shapes, demonstrate that the shape is undefined for a vanishing density gradient, explain how we derived the criterion for the shape sensitivity, and discuss the role of the convergence criterion.

\subsection{Measuring shapes} \label{sec:shapes_measure}

In the literature various methods have been applied to measure shapes \citep[e.g.][]{Allgood_2006, Bett_2007, Warnick_2008, Bett_2012, Schneider_2012, Peter_2013, Robertson_2019, Banerjee_2020, Chua_2021a, Sameie_2018, Fischer_2021b, Harvey_2021, Vargya_2021}.
An overview of different methods is given by \cite{Zemp_2011}. We used their two recommended methods.
Both are iterative procedures where we selected particles in a given volume and inferred their shape from the mass tensor:
\begin{equation} \label{eq:mass_tensor}
    \mathbf{M}_{ij} = \sum_k m_k r_{k,i} r_{k,j}
.\end{equation}
Here, $k$ denotes a particle and $i$, $j$ are the coordinate indices.
The axis ratios follow from the eigenvalues ($\lambda_1\geq\lambda_2\geq\lambda_3$): $q = \sqrt{\lambda_2/\lambda_1}$ and $s=\sqrt{\lambda_3/\lambda_1}$.
We used them to compute the new shape of the selection volume and the eigenvectors to compute its orientation.
The mass tensor was again computed from the particles in the new selection volume.
We continued this procedure until the axis ratios of the selection volume converged against the one calculated from the mass tensor.

The two methods considered in this work selected particles within ellipsoids and ellipsoidal shells (homoeoids).
The volume of an ellipsoid is given by $V = 4/3 \pi \, a \, b \, c$, with the semi-axes $a\geq b\geq c$.
The radius that corresponds to the same volume of a sphere is $r^3 = a \, b \, c$.
Their ratios are $q=b/a$ and $s=c/a$.
To select the particles within a volume, we introduced
\begin{equation}
    \Tilde{r}^2_\mathrm{ell} := x^2 \, (q \, s)^{2/3} + y^2 \, q^{-4/3} \, s^{2/3} + z^2 \, q^{2/3} \, s^{-4/3} \,,
\end{equation}
which ensured that the volume of the ellipsoids remains constant.
Here, $x$, $y$, and $z$ denote the spatial coordinates.
If $\Tilde{r}^2_\mathrm{ell}$ was smaller than $r$, the particles were selected for the ellipsoids.
In the case of the ellipsoidal shells, there also exists a lower limit on $\Tilde{r}^2_\mathrm{ell}$.
In practice, we used the same implementation as previously used in \cite{Fischer_2022}.

\subsection{Density gradient} \label{sec:shape_gradient}
Here, we briefly demonstrate the relevance of the density gradients for halo shapes by showing that the shape is undefined if the gradient vanishes.
To do so, we assumed a constant density $\rho$ and tried to measure its shape.
The continuous version of Eq.~\ref{eq:mass_tensor} is given by
\begin{equation} \label{eq:mass_tensor2}
    \mathbf{M}_{ij} = \int_V \rho \, r_{i} \, r_{j} \, \mathrm{d}V \,.
\end{equation}
For measuring the shape within an ellipsoidal shell with the semi-axes $a$, $b$, and $c$, we introduced new coordinates,
\begin{equation}
    \mathbf{r} = p \left(\begin{matrix} a \, \sin \theta \, \cos \varphi \\ b \, \sin \theta \, \sin \varphi \\ c \, \cos \theta \end{matrix}\right) \,,
\end{equation}
with values of $0 \leq p \leq 1$ for positions within the ellipsoid.
Now, we can rewrite Eq.~\ref{eq:mass_tensor2} as
\begin{equation}
    \mathbf{M}_{ij} = \rho \int_{p_1}^{p_2} \int_0^\pi \int_0^{2\pi} r_{i} \, r_{j} \, a \, b \, c \, p^2 \, \sin \theta \, \mathrm{d}\varphi \, \mathrm{d}\theta \, \mathrm{d}p \,.
\end{equation}
The edges of the elliptical shell are given by $p_1$ and $p_2$.
By integration, we found that all off-diagonal elements of $\mathbf{M}$ are zero.
In consequence, the eigenvalues are $\mathbf{M}_{11} = g \, a^2$, $\mathbf{M}_{22} = g \, b^2$, and $\mathbf{M}_{33} = g \, c^2$ with $g:=4 \pi \, 15^{-1} \rho \, a \, b \, c \, (p_2^5 - p_1^5)$.
Hence we obtained the shape of the integration volume, rendering the shape of the mass distribution undefined.
It is therefore independent of the method used to determine the shape.
An iterative procedure such as the one described in Sect.~\ref{sec:shapes_measure} could not converge against any meaningful value.
In consequence, the density gradient is essential to the definition of the shape of a mass distribution.

\subsection{Criterion for the shape sensitivity} \label{sec:shape_crit}

With the explanation above, it becomes clear that the density gradient is a determining factor for the sensitivity of the shape measurement.
If it is low, the discretisation error may dominate and the shape can no longer be calculated reliably.
In the case of measuring the shape within ellipsoidal shells, one can calculate the sensitivity as follows.
The discretisation noise can be estimated as $\sqrt{N_i}$, where $N_i$ is the number of particles in a shell $i$.
Hence, the relative error is $N_i^{-1/2}$.
Given the number density of the shell, $n_i$, its error is then $n_i \, N_i^{-1/2}$.
We can relate this to the density gradient by simply computing the finite difference in the number density of the neighbouring shells.
Thus, a condition for the shape measurement to be sensitive can be as follows:
\begin{equation}
    |n_{i-1} - n_{i+1}| > n_i \, N_i^{-1/2} \, .
\end{equation}
From this, we derived $\xi$ as a measure of the sensitivity, which should be larger than unity,
\begin{equation} \label{eq:shape_critS}
    \xi_i \equiv \frac{V_i}{\sqrt{N_i}} \left| \frac{N_{i-1}}{V_{i-1}} - \frac{N_{i+1}}{V_{i+1}}\right| > 1,\end{equation}
where $V_i$ denotes the volume of shell $i$, with $n_i = N_i / V_i$.
Alternatively one may express the condition as
\begin{equation}
    \Delta x_i \, \left| \frac{\mathrm{d}n}{\mathrm{d}x}\Big|_i \right| > n_i \, N^{-1/2}_i \,,
    \label{eq:alternative_criterion}
\end{equation}
where $\Delta x_i$ denotes the thickness of the shell.

We want to point out that the shape sensitivity, $\xi_i$, also depends on the binning.
From Eq.~\ref{eq:alternative_criterion} it can be seen that the sensitivity decreases for smaller radial ellipsoidal bins.
Hence, there exists a tradeoff between the radial resolution and the accuracy of the measured shapes.

When not using shells, but all particles included in an ellipsoid, a similar criterion can be obtained.
To be able to estimate the noise due to the numerical discretisation, we introduced ``pseudo-shells''.
The idea is that the outermost particles of the ellipsoid are the main contributors to the mass tensor and they therefore lie within a pseudo-shell with thickness $\Delta x_i$\,  thus allowing us to estimate the noise.
We introduced a parameter $\gamma = r_2/r_1$ to define the relative thickness of the pseudo-shell, with $r_1 > r_2$ being the radius of the ellipsoid and the inner pseudo-shell radius, respectively.
Thus, the thickness of the pseudo-shell is defined as $\Delta x_i := r_1 - r_2$.

The thickness $\Delta x_i$ from Eq.~\ref{eq:alternative_criterion} then was found to be $\Delta x_i = x_i (1-\gamma)$, with the ellipsoid radius $x_i$.
Now we can rewrite the condition as follows:
\begin{equation} \label{eq:shape_critE}
    \xi_i \equiv \sqrt{\frac{4 \pi}{3 n_i}} \left| \frac{\mathrm {d}n}{\mathrm{d}x}\Big|_i \right| \, x_i^{5/2} \, \alpha > 1 \,.
\end{equation}
For this, we introduced $\alpha:=(1-\gamma)\sqrt{1-\gamma^3}$ and assumed it to be constant\footnote{Strictly speaking, $\alpha$ depends on the density gradient. For steep gradients, it would be larger than for flat ones as in the first case a larger contribution to the mass tensor stems from smaller radii. Nevertheless, we set $\alpha$ to a constant value to simplify the calculations.}.
In the following we use $\alpha = 0.5$, because we found it to provide similar results for the shape sensitivity profile compared to the method based on shells (readers can compare the example haloes shown in Sect.~\ref{sec:results} and Appendix~\ref{sec:shape_shells}).
In practice, we computed the density gradient, $\mathrm{d}n/\mathrm{d}x$, by taking the differences in mass, volume, and $\tilde{r}_\mathrm{ell}$ between two consecutive ellipsoids.

According to the derivation above, $\xi>1$ is required only.
However, in practice we recommend using a more stringent criterion, for example, $\xi>10$, to include a safety margin.
We found a value of 1 to be too low; instead, a value of 10 was motivated by our findings in Sect.~\ref{sec:results} and Appendix~\ref{sec:shape_shells}.
Here one can see that for lower values, the shapes are not reliable, as it does not make sense that the shape profile of the frequently self-interacting dark matter (fSIDM) haloes becomes more elliptical and noisy in the central region.
From here on we refer to $\xi$ as the definition of the shape sensitivity.

\subsection{Convergence criterion}

The convergence criterion plays a major role.
If the typical fluctuations in the axis ratios arising from the discrete nature of the $N$-body representation are larger than the convergence criterion, the shapes can become more and more elliptical with each iteration.
To solve this problem, the convergence criterion could be chosen less strictly such that it is larger than the discretisation noise.
However, this would come at the cost of being biased towards the initial chosen shape, which is typically a sphere.
We did not follow this approach but chose a strict convergence criterion in the following and required that the inferred axis ratios did not vary by more than $0.1\%$ per iteration.
If we did not converge within 50 iterations, we stopped and took the last shape found. 
In practice, this occurs only if a small number of particles is used for the tensor calculation and thus this affects the shape at small radii only.

\section{Simulated haloes} \label{sec:results}

\begin{figure*}
    \centering
    \includegraphics[width=\textwidth]{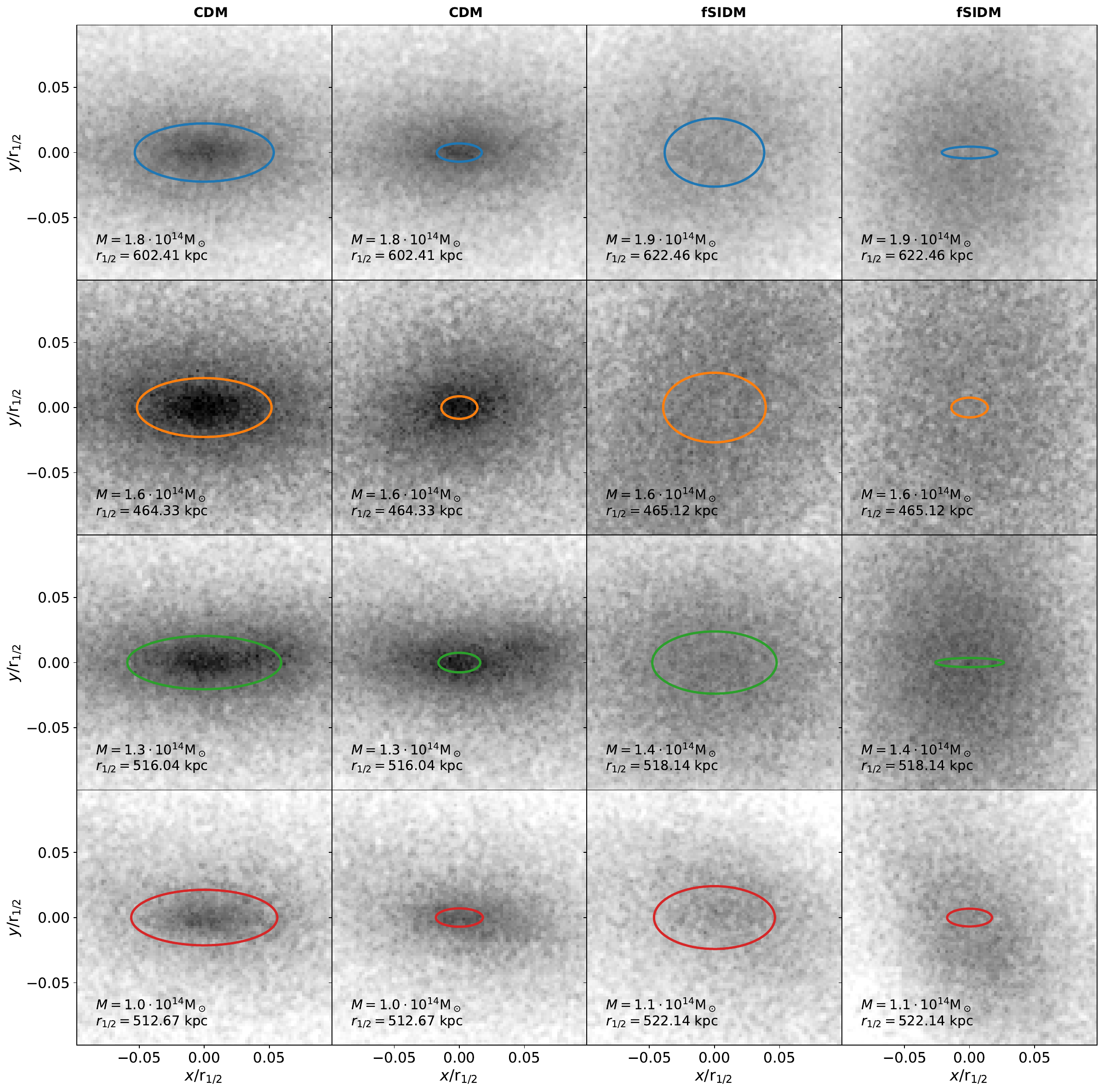}
    \caption{
    DM surface density is shown for the primary subhaloes simulated with CDM (the two left columns) and fSIDM with $\sigma_\mathrm{\Tilde{T}}/m = 0.1 \, \mathrm{cm}^2 \, \mathrm{g}^{-1}$ (the two right columns).
    We note that the colour scaling is logarithmic.
    The shapes were measured within a volume of $V = 4/3 \, \pi \, (0.032 \, r_\mathrm{1/2})^3$ (odd columns) and $V = 4/3 \, \pi \, (0.0105 \, r_\mathrm{1/2})^3$ (even columns).
    They are illustrated with coloured ellipses.
    The haloes are shown face-on according to the measured shapes.
    We matched the haloes across the simulation, where matched haloes are shown in the same row and the same colour is used for the ellipses.}
    \label{fig:map}
\end{figure*}

\begin{figure*}
    \centering
    \includegraphics[width=\columnwidth]{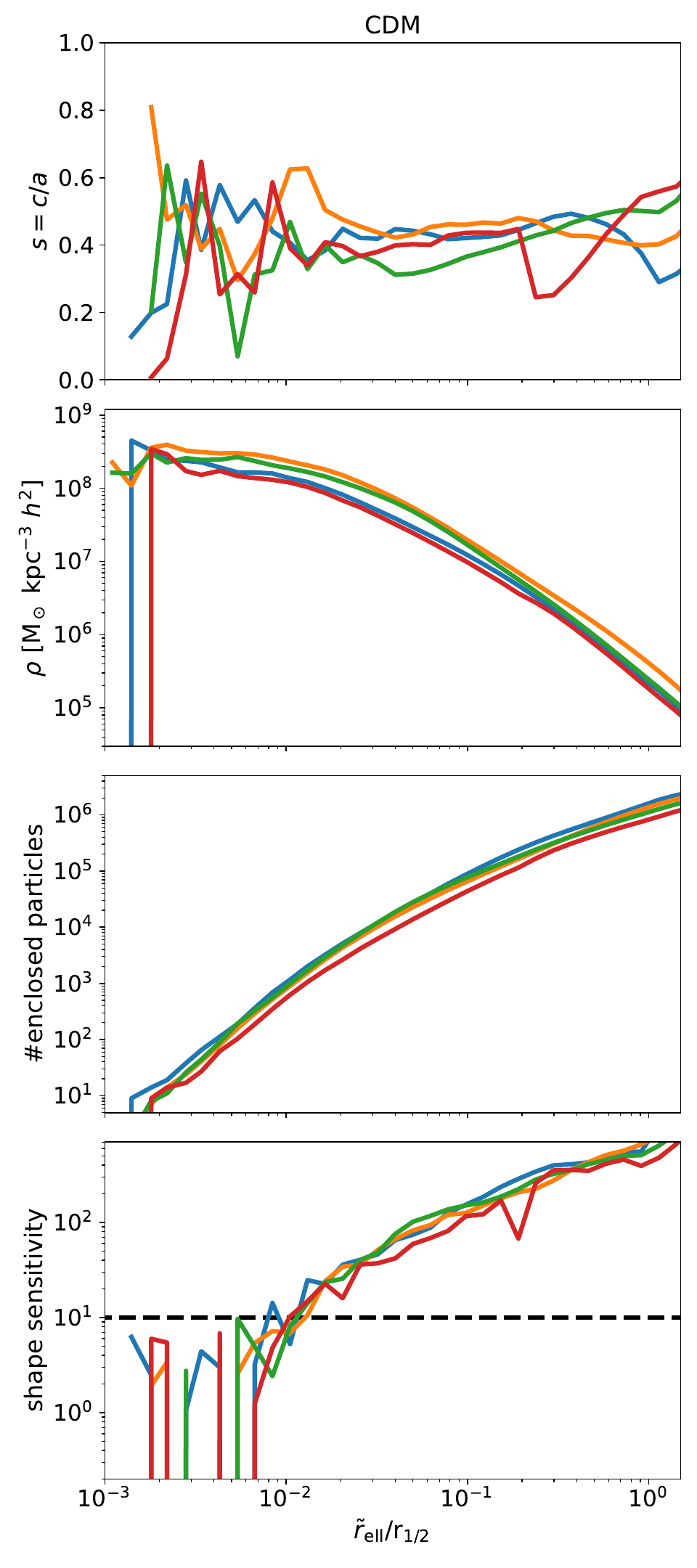}
    \includegraphics[width=\columnwidth]{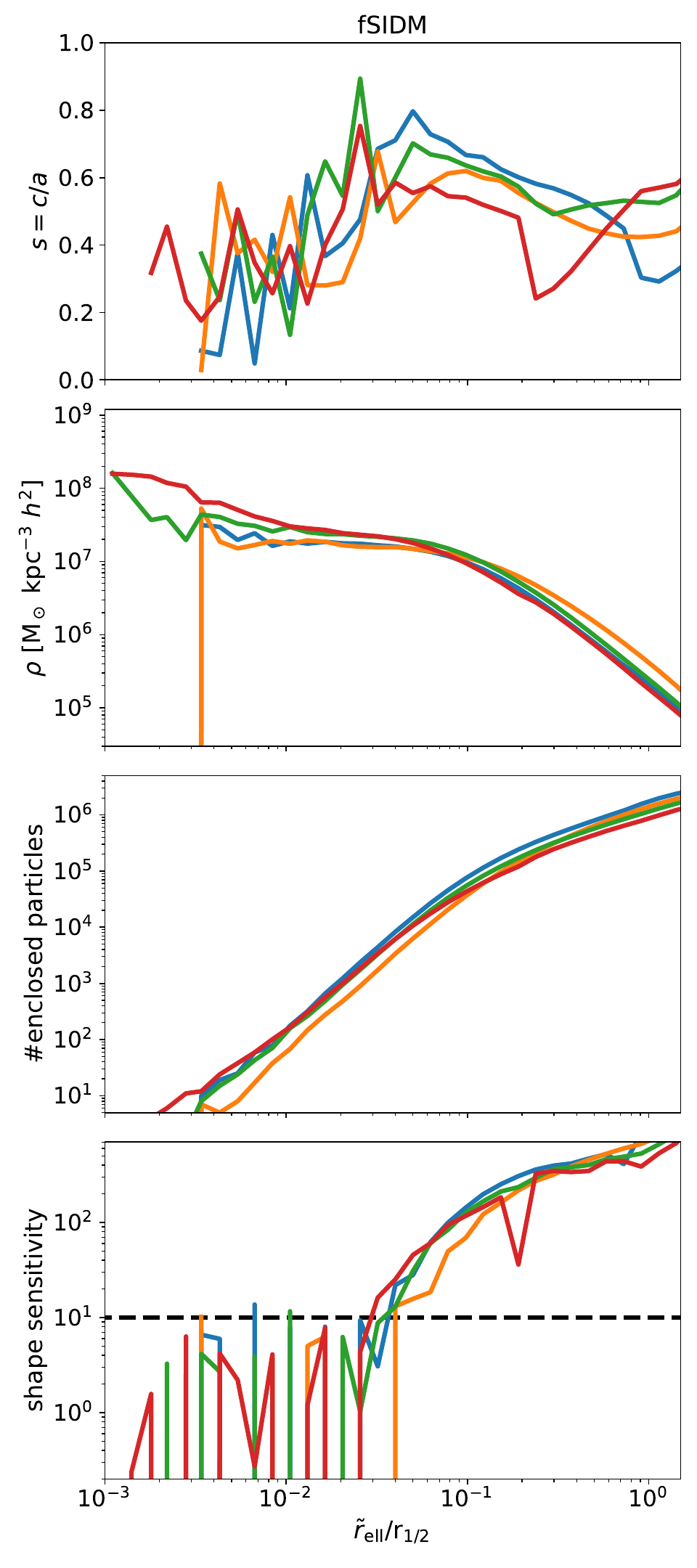}
    \caption{Various quantities for the CDM (right-hand side) and fSIDM (left-hand side) haloes (the colours are the same as in Fig.~\ref{fig:map}) are shown as a function of $\Tilde{r}_\mathrm{ell}$.
    First row: Shape profile as the ratio of the major and minor axis is shown. The mass tensor was measured within ellipsoids. Second row: Density profile is shown as inferred from the ellipsoids used for the shape measurement by taking the differences in mass and volume between two consecutive ellipsoids. Third row: Number of particles enclosed in the ellipsoid is shown. Fourth row: Shape sensitivity, $\xi$, is shown according to Eq.~\ref{eq:shape_critE}. Here the vertical line indicates the criterion above which the shape was measured accurately, $\xi = 10$.
    }
    \label{fig:profile_1}
\end{figure*}

In this section, we explain how we measured the shape of simulated DM haloes and investigated their accuracy.
We used the four most massive haloes of the full box cosmological simulations presented in \cite{Fischer_2022}. 
They have total masses of about ${\sim} 10^{14} \mathrm{M_\odot}$ and are therefore considered low-mass galaxy clusters.
We used these haloes as they are the best-resolved ones with about $N \sim 2.3 \times 10^{6}$ particles.
The maximal physical Plummer-equivalent gravitational softening length is $1.4\,\mathrm{kpc}$.
The same analysis as in \cite{Fischer_2022} to identify the substructure and compute corresponding quantities such as the halo mass and the virial radius was employed, that is, we used \textsc{SubFind} \citep{Springel_2001, Dolag_2009}.
We computed the total mass, $M$, as the sum of the gravitationally bound particles as identified by \textsc{SubFind}.
To quantify the spatial extent of the particles, we used the half-mass radius, $r_{1/2}$, that is the radius that contains half of the gravitationally bound mass around the gravitationally most-bound particle.
Moreover, we used the simulations with the highest resolution ($N = 576^3$, $m_\mathrm{DM} = 4.37 \times 10^7 \, \mathrm{M_\odot} \, h^{-1}$) for collisionless cold dark matter (CDM) and fSIDM with a momentum transfer cross-section of $\sigma_\mathrm{\Tilde{T}}/m = 0.1 \, \mathrm{cm}^2 \, \mathrm{g}^{-1}$ \citep{Kahlhoefer_2014, Fischer_2021a}.
For further details, readers can refer to \cite{Fischer_2022}.

In Fig.~\ref{fig:map} we show the DM surface density in the central region of the four most massive haloes to illustrate the problem at hand.
Only the particles belonging to the primary subhalo as identified by \textsc{SubFind} were used, and as a result of which we avoided including satellite substructure.
We measured the shape within a volume of $V = 4/3 \, \pi \, (0.032 \, r_\mathrm{1/2})^3$ and $V = 4/3 \, \pi \, (0.0105 \, r_\mathrm{1/2})^3$, indicated by the coloured ellipses.
The exact volumes are arbitrary\footnote{We simply used two of the elliptical volumes that we computed for Fig.~\ref{fig:profile_1}.}, but they were selected with the aim to illustrate that the shape measurement can fail within the very central region while it is working at larger distances.
The primary subhaloes are aligned according to this shape measurement and are shown face-on.
We note that we show the very inner region of the haloes.
The half-mass radius of the haloes is roughly about ${\sim}35\%$ of the virial radius.
Many studies have focussed on the halo shapes at larger distances.
But there has also been interest in the inner region of haloes \citep[e.g.][]{de_Nicola_2022}.
We find that the CDM haloes (left-hand side) are more elliptical than the fSIDM haloes (right-hand side).
By eye, the inferred shapes seem to roughly match the underlying DM distribution for the odd columns.
This is in contrast to the even columns where the same thing was done, but we measured the shape at smaller distances to the centre.
For the CDM haloes (left-hand side), they seem to agree with the surface density plots, but they do not match for fSIDM haloes (right-hand side).
For the latter case, the volume used for the shape measurement lies within the density core that formed due to the DM self-interactions.
According to Sect.~\ref{sec:introduction} we expect that the vanishing density gradient is a limiting factor here.
So far, we have illustrated that the shape measurement can go wrong for a halo when additional physics lead to a density core.
In fact, the sensitivity of the shape depends on both the numerical resolution and the density gradient.
While these two effects cannot be disentangled, their combined effect can be measured with a shape sensitivity criterion (see Sect.~\ref{sec:shape_crit}).
In the following, we investigate this quantitatively.

In Fig.~\ref{fig:profile_1} we show the shape and the density profiles as well as the number of particles enclosed within the ellipsoids and the shape sensitivity as a function of $\tilde{r}_\mathrm{ell}$.
In contrast to Fig.~\ref{fig:map}, we now quantitatively assess the quality of the measured shapes.
In Appendix~\ref{sec:shape_shells} we explain how we carried out the same analysis, but measured the shapes in shells.
We can clearly see that the density core present in the fSIDM haloes (right-hand side) affects the accuracy at which the shapes can be measured compared to the CDM (left-hand side).
At small distances, the inferred shape from the fSIDM haloes (upper panel, right-hand side) can be even more elliptical than the CDM one (upper panel, left-hand side).
This is non-physical and an artefact arising from the low shape sensitivity.
It is insufficient to explain this by a lower number of enclosed particles as the shape measure becomes unreliable at particle numbers for which the CDM shapes are still reasonable.
This can be seen by looking at the number of enclosed particles for the distance at which the measured shapes (top row) become fairly noisy.
For the CDM haloes, the shape starts to become fairly noisy for $\lesssim 10^3$ particles, but the fSIDM halo shapes suffer from substantial noise already for $\lesssim 10^4$ particles.
For both the ellipsoid and the ellipsoidal shell methods, a shape sensitivity of $\xi \geq 10$ is found to be a good criterion for obtaining robust shape measurements, as can be seen from the top and bottom rows of Figs.~\ref{fig:profile_1} and~\ref{fig:profile_2}.

We note that it can also be the case that at large radii the shape cannot be measured accurately.
This could be due to substructure; an example would be the halo indicated in red.
As we only selected particles that belong to the primary subhalo, this effect has been suppressed however.

\section{Discussion} \label{sec:discussion}

In this section, we discuss the implications of our results regarding the sensitivity of halo shape measurements.
In cosmological simulations, a convergence radius according to \cite{Power_2003} is often computed.
\cite{Stadel_2009} noticed that the convergence radius for the shape in DM-only simulations of CDM is roughly three times larger than the one of the density profile.
We found that additional physical processes can prohibit safely determining the shape beyond a certain radius defined by the numerical resolution.

In contrast, the accuracy of measured shapes does not purely depend on the numerical properties, but also on the underlying physical matter distribution.
Implying that a different criterion, such as the one derived in Sect.~\ref{sec:shape_crit}, is needed, taking the density gradient into account.
Although we have studied haloes with a mass of ${\sim} 10^{14} \, \mathrm{M_\odot}$, these findings are independent of the halo mass.

It is crucial to be aware of this when comparing simulation results to observations.
For example, the deprojection algorithm of \cite{de_Nicola_2020} allows observed galaxies to be inferred and it can be used for comparison \citep{de_Nicola_2022}.
Interestingly, \cite{de_Nicola_2022} found the shape in the most inner region of the brightest cluster galaxies to become more elliptical, which is in contrast to previous findings where the baryons were found to make haloes more spherical \citep[e.g.][]{Cataldi_2021}.
Such results have to be taken with caution if the accuracy of the simulated shapes is unclear.
This is even true for pure theoretical studies that investigate the impact of feedback on the halo shape for example \citep[e.g.][]{Chua_2022}.
An approach different from computing 3D shapes would be to create mock observations from the simulation data and to use them for comparison as was done by \cite{Harvey_2020} or to investigate mock weak lensing data \citep{Robertson_2022}.

We found that density cores are prohibitive for reliable shape measurements.
The ones we studied were formed by DM self-interactions, but also other mechanisms such as baryonic processes, for example, supernovae \citep[e.g.][]{Read_2005} or black hole feedback \citep[e.g.][]{Martizzi_2013}, can reduce the density gradient in the halo centre.
Whenever a density core is present, it undermines the ability to measure meaningful shapes.

Depending on the density profile, shapes cannot be measured at arbitrarily small radii.
Increasing the resolution of the simulation would hardly improve the situation if density cores continue to be present.
Instead, we suggest measuring shapes only beyond the radius of the density core and verifying their reliability with our shape sensitivity criterion.

\section{Conclusion} \label{sec:conclusion}

In this  paper, we have introduced a criterion to estimate the sensitivity of shape measurements from $N$-body simulation data and measured the shapes of various haloes to verify the criterion.
Our main conclusions are as follows:
\begin{itemize}
    \item The accuracy of measured shapes does not only depend on the numerical resolution, but also on the density gradient.
    \item Density cores lead to problematic measurements of shapes in the central region of haloes.
    \item Care must be taken as to the quality of measured shapes, which can be inferred from the criterion in Eq.~\ref{eq:shape_critS} (Eq.~\ref{eq:shape_critE}) when the shape is measured employing a mass tensor computed from particles in elliptical shells (ellipsoids).
\end{itemize}
Future studies can improve the reliability of their measured shapes by employing the presented shape sensitivity criterion. In this way it is possible to ensure physically meaningful shapes are determined.

\begin{acknowledgements}
We thank Marcus Brüggen and Klaus Dolag for comments and Stefano de Nicola, Roberto P.\ Saglia and Jens Thomas for inspiring discussion.
We also thank the anonymous referees for comments that improved the paper.
LMV acknowledges support by the COMPLEX project from the European Research Council (ERC) under the European Union’s Horizon 2020 research and innovation program grant agreement ERC-2019-AdG 882679.
This work is funded by the Deutsche Forschungsgemeinschaft (DFG, German Research Foundation) under Germany's Excellence Strategy -- EXC 2121 ``Quantum Universe'' --  390833306, Germany’s Excellence Strategy -- EXC-2094 ``Origins'' -- 390783311.

Software:
NumPy \citep{NumPy},
Matplotlib \citep{Matplotlib}.
\end{acknowledgements}

\bibliographystyle{style/aa}
\bibliography{bib.bib}

\begin{appendix}
\section{Shape measured in shells} \label{sec:shape_shells}

\begin{figure*}
    \centering
    \includegraphics[width=\columnwidth]{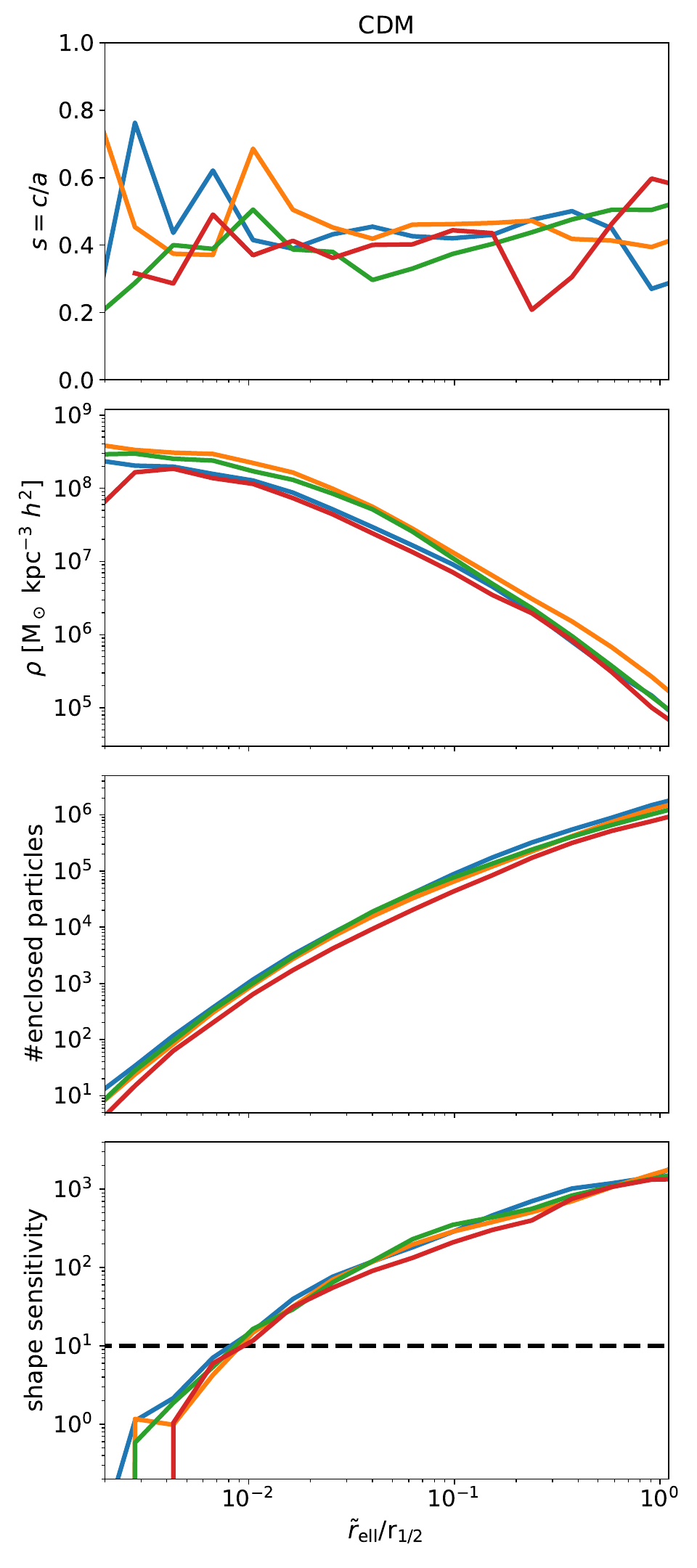}
    \includegraphics[width=\columnwidth]{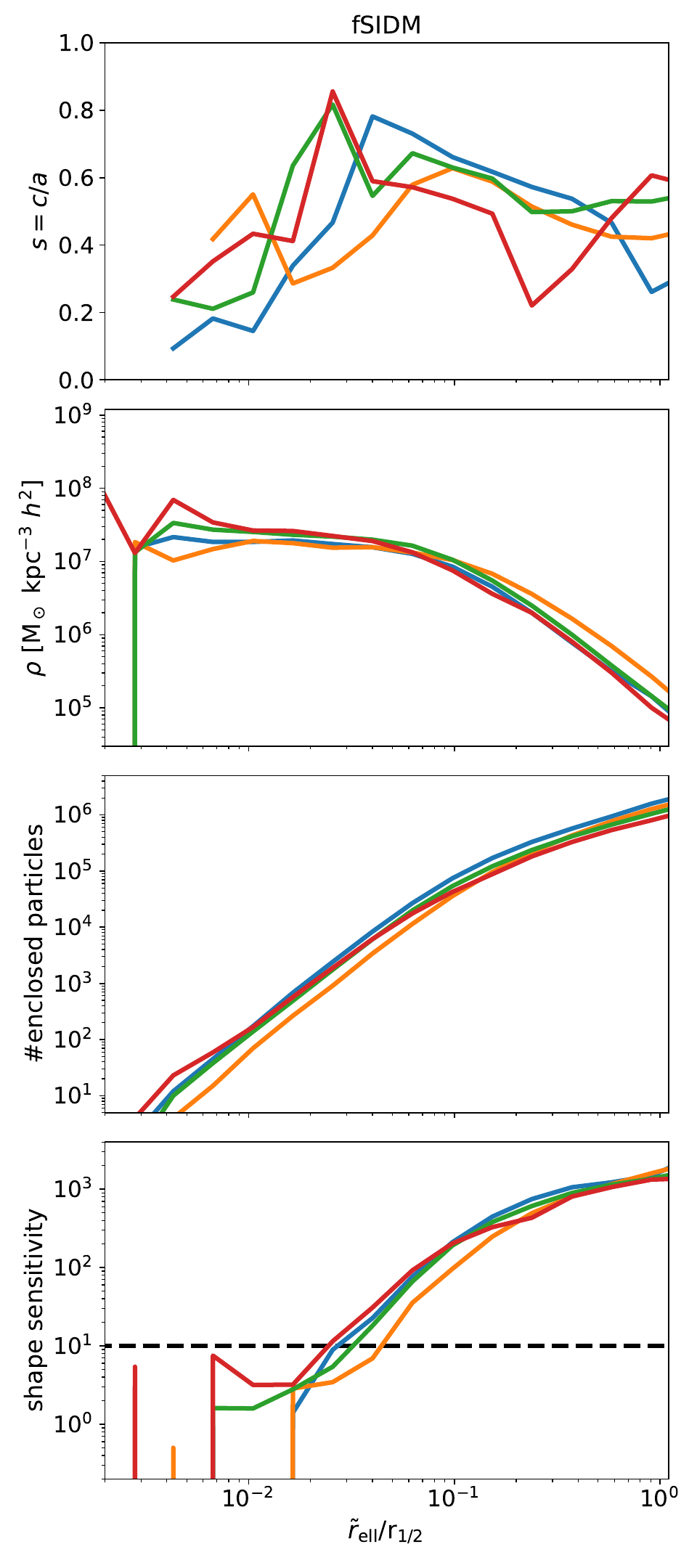}
    \caption{Same plot as in Fig.~\ref{fig:profile_1}, but the shapes were computed using only particles within shells and not all particles are enclosed in an ellipsoid. In consequence, Eq.~\ref{eq:shape_critS} was used to compute the shape sensitivity.
    }
    \label{fig:profile_2}
\end{figure*}

For this appendix, we computed the shapes based on particles within shells.
Consequently we used Eq.~\ref{eq:shape_critS} to compute the shape sensitivity.
Analogously to Fig.~\ref{fig:profile_1}, we display the measured shapes and their sensitivity in Fig.~\ref{fig:profile_2}.
Qualitatively we found the same thing as before for the particles enclosed in ellipsoids.

\end{appendix}

\end{document}

%% file: style/settings.tex
\makeatletter
\renewcommand*\aa@pageof{, page \thepage{} of \pageref*{LastPage}}
\makeatother

%% file: style/packages.tex
\usepackage{bm} 
\usepackage{textgreek} 

\usepackage[capitalize,noabbrev]{cleveref} 
\creflabelformat{equation}{#2#1#3} 
\crefname{enumi}{item}{items} 

\usepackage{siunitx} 
\DeclareSIUnit[number-unit-product = ]\percent{\char`\%} 

\usepackage{csquotes} 
\usepackage[normalem]{ulem}

%% file: style/definitions.tex
\DeclareSIUnit\parsec{pc}
\DeclareSIUnit\dex{dex}
\DeclareSIUnit\h{\mathnormal{h}}
\DeclareSIUnit\year{yr}
\DeclareSIUnit\years{yrs}
\DeclareSIUnit\arcsec{arcsec}
\DeclareSIUnit\arcmin{arcmin}
\DeclareSIUnit\Msun{M_\odot}
\DeclareSIUnit\Rsun{R_\odot}
\DeclareSIUnit\Lsun{L_\odot}
\DeclareSIUnit\Rvir{\mathnormal{R}_\mathrm{vir}}
\DeclareSIUnit\Rhalf{\mathnormal{R}_{1/2}}
\DeclareSIUnit\erg{erg}
\DeclareSIUnit\angstrom{\text{Å}}

\newcommand*{\Msun}{\ensuremath{\mathrm{M}_\odot}} 
\newcommand*{\Rsun}{\ensuremath{\mathrm{R}_\odot}} 
\newcommand*{\Lsun}{\ensuremath{\mathrm{L}_\odot}} 
\newcommand*{\Rvir}{\ensuremath{R_\mathrm{vir}}} 
\newcommand*{\Rhalf}{\ensuremath{R_{1/2}}} 


%% file: main.bbl
\begin{thebibliography}{55}
\expandafter\ifx\csname natexlab\endcsname\relax\def\natexlab#1{#1}\fi

\bibitem[{Abadi {et~al.}(2010)Abadi, Navarro, Fardal, Babul, \&
  Steinmetz}]{Abadi_2010}
Abadi, M.~G., Navarro, J.~F., Fardal, M., Babul, A., \& Steinmetz, M. 2010,
  \mnras, 407, 435

\bibitem[{Allgood {et~al.}(2006)Allgood, Flores, Primack, Kravtsov, Wechsler,
  Faltenbacher, \& Bullock}]{Allgood_2006}
Allgood, B., Flores, R.~A., Primack, J.~R., {et~al.} 2006, \mnras, 367, 1781

\bibitem[{Banerjee {et~al.}(2020)Banerjee, Adhikari, Dalal, More, \&
  Kravtsov}]{Banerjee_2020}
Banerjee, A., Adhikari, S., Dalal, N., More, S., \& Kravtsov, A. 2020, \jcap,
  2020, 024

\bibitem[{Bett(2012)}]{Bett_2012}
Bett, P. 2012, \mnras, 420, 3303

\bibitem[{Bett {et~al.}(2007)Bett, Eke, Frenk, Jenkins, Helly, \&
  Navarro}]{Bett_2007}
Bett, P., Eke, V., Frenk, C.~S., {et~al.} 2007, \mnras, 376, 215

\bibitem[{Bonamigo {et~al.}(2015)Bonamigo, Despali, Limousin, Angulo, Giocoli,
  \& Soucail}]{Bonamigo_2015}
Bonamigo, M., Despali, G., Limousin, M., {et~al.} 2015, \mnras, 449, 3171

\bibitem[{Bryan {et~al.}(2013)Bryan, Kay, Duffy, Schaye, Vecchia, \&
  Booth}]{Bryan_2013}
Bryan, S.~E., Kay, S.~T., Duffy, A.~R., {et~al.} 2013, \mnras, 429, 3316

\bibitem[{Butsky {et~al.}(2016)Butsky, Macci{\`{o}}, Dutton, Wang, Obreja,
  Stinson, Penzo, Kang, Keller, \& Wadsley}]{Butsky_2016}
Butsky, I., Macci{\`{o}}, A.~V., Dutton, A.~A., {et~al.} 2016, \mnras, 462, 663

\bibitem[{Cataldi {et~al.}(2021)Cataldi, Pedrosa, Tissera, \&
  Artale}]{Cataldi_2021}
Cataldi, P., Pedrosa, S.~E., Tissera, P.~B., \& Artale, M.~C. 2021, \mnras,
  501, 5679

\bibitem[{Chua {et~al.}(2021)Chua, Dibert, Vogelsberger, \&
  Zavala}]{Chua_2021a}
Chua, K. T.~E., Dibert, K., Vogelsberger, M., \& Zavala, J. 2021, \mnras, 500,
  1531

\bibitem[{Chua {et~al.}(2019)Chua, Pillepich, Vogelsberger, \&
  Hernquist}]{Chua_2019}
Chua, K. T.~E., Pillepich, A., Vogelsberger, M., \& Hernquist, L. 2019, \mnras,
  484, 476

\bibitem[{Chua {et~al.}(2022)Chua, Vogelsberger, Pillepich, \&
  Hernquist}]{Chua_2022}
Chua, K. T.~E., Vogelsberger, M., Pillepich, A., \& Hernquist, L. 2022, \mnras,
  515, 2681

\bibitem[{de~Nicola {et~al.}(2022)de~Nicola, Saglia, Thomas, Pulsoni, Kluge,
  Bender, Valenzuela, \& Remus}]{de_Nicola_2022}
de~Nicola, S., Saglia, R.~P., Thomas, J., {et~al.} 2022, \apj, 933, 215

\bibitem[{Despali {et~al.}(2014)Despali, Giocoli, \& Tormen}]{Despali_2014}
Despali, G., Giocoli, C., \& Tormen, G. 2014, \mnras, 443, 3208

\bibitem[{Despali {et~al.}(2013)Despali, Tormen, \& Sheth}]{Despali_2013}
Despali, G., Tormen, G., \& Sheth, R.~K. 2013, \mnras, 431, 1143

\bibitem[{Despali {et~al.}(2022)Despali, Walls, Vegetti, Sparre, Vogelsberger,
  \& Zavala}]{Despali_2022}
Despali, G., Walls, L.~G., Vegetti, S., {et~al.} 2022, \mnras, 516, 4543

\bibitem[{de Nicola {et~al.}(2020)de Nicola, Saglia, Thomas, Dehnen, \&
  Bender}]{de_Nicola_2020}
de Nicola, S., Saglia, R.~P., Thomas, J., Dehnen, W., \& Bender, R. 2020,
  \mnras, 496, 3076

\bibitem[{Dolag {et~al.}(2009)Dolag, Borgani, Murante, \&
  Springel}]{Dolag_2009}
Dolag, K., Borgani, S., Murante, G., \& Springel, V. 2009, \mnras, 399, 497

\bibitem[{Emami {et~al.}(2021{\natexlab{a}})Emami, Genel, Hernquist, Alcock,
  Bose, Weinberger, Vogelsberger, Marinacci, Loeb, Torrey, \&
  Forbes}]{Emami_2021a}
Emami, R., Genel, S., Hernquist, L., {et~al.} 2021{\natexlab{a}}, \apj, 913, 36

\bibitem[{Emami {et~al.}(2021{\natexlab{b}})Emami, Hernquist, Alcock, Genel,
  Bose, Weinberger, Vogelsberger, Shen, Speagle, Marinacci, Forbes, \&
  Torrey}]{Emami_2021b}
Emami, R., Hernquist, L., Alcock, C., {et~al.} 2021{\natexlab{b}}, \apj, 918, 7

\bibitem[{Fischer {et~al.}(2021{\natexlab{a}})Fischer, Brüggen,
  Schmidt-Hoberg, Dolag, Kahlhoefer, Ragagnin, \& Robertson}]{Fischer_2021a}
Fischer, M.~S., Brüggen, M., Schmidt-Hoberg, K., {et~al.} 2021{\natexlab{a}},
  \mnras, 505, 851

\bibitem[{Fischer {et~al.}(2022)Fischer, Brüggen, Schmidt-Hoberg, Dolag,
  Kahlhoefer, Ragagnin, \& Robertson}]{Fischer_2022}
Fischer, M.~S., Brüggen, M., Schmidt-Hoberg, K., {et~al.} 2022, \mnras, 516,
  1923

\bibitem[{Fischer {et~al.}(2021{\natexlab{b}})Fischer, Brüggen,
  Schmidt-Hoberg, Dolag, Ragagnin, \& Robertson}]{Fischer_2021b}
Fischer, M.~S., Brüggen, M., Schmidt-Hoberg, K., {et~al.} 2021{\natexlab{b}},
  \mnras, 510, 4080

\bibitem[{Harris {et~al.}(2020)Harris, Millman, van~der Walt, Gommers,
  Virtanen, Cournapeau, Wieser, Taylor, Berg, Smith, Kern, Picus, Hoyer, van
  Kerkwijk, Brett, Haldane, Fern\'{a}ndez~del R\'{i}o, Wiebe, Peterson,
  G\'{e}rard-Marchant, Sheppard, Reddy, Weckesser, Abbasi, Gohlke, \&
  Oliphant}]{NumPy}
Harris, C.~R., Millman, K.~J., van~der Walt, S.~J., {et~al.} 2020, Nature, 585,
  357

\bibitem[{Harvey {et~al.}(2021)Harvey, Chisari, Robertson, \&
  McCarthy}]{Harvey_2021}
Harvey, D., Chisari, N.~E., Robertson, A., \& McCarthy, I.~G. 2021, \mnras,
  506, 441–451

\bibitem[{Harvey {et~al.}(2020)Harvey, Robertson, Tam, Jauzac, Massey, Rhodes,
  \& McCarthy}]{Harvey_2020}
Harvey, D., Robertson, A., Tam, S.-I., {et~al.} 2020, \mnras, 500, 2627

\bibitem[{{Hellwing} {et~al.}(2021){Hellwing}, {Cautun}, {van de Weygaert}, \&
  {Jones}}]{Hellwing_2021}
{Hellwing}, W.~A., {Cautun}, M., {van de Weygaert}, R., \& {Jones}, B.~T. 2021,
  \prd, 103, 063517

\bibitem[{{Hunter}(2007)}]{Matplotlib}
{Hunter}, J.~D. 2007, Computing in Science Engineering, 9, 90

\bibitem[{Jing \& Suto(2002)}]{Jing_2002}
Jing, Y.~P. \& Suto, Y. 2002, \apj, 574, 538

\bibitem[{{Kahlhoefer} {et~al.}(2014){Kahlhoefer}, {Schmidt-Hoberg},
  {Frandsen}, \& {Sarkar}}]{Kahlhoefer_2014}
{Kahlhoefer}, F., {Schmidt-Hoberg}, K., {Frandsen}, M.~T., \& {Sarkar}, S.
  2014, \mnras, 437, 2865

\bibitem[{{Katz}(1991)}]{Katz_1991}
{Katz}, N. 1991, \apj, 368, 325

\bibitem[{Kazantzidis {et~al.}(2010)Kazantzidis, Abadi, \&
  Navarro}]{Kazantzidis_2010}
Kazantzidis, S., Abadi, M.~G., \& Navarro, J.~F. 2010, \apj, 720, L62

\bibitem[{Martizzi {et~al.}(2013)Martizzi, Teyssier, \& Moore}]{Martizzi_2013}
Martizzi, D., Teyssier, R., \& Moore, B. 2013, \mnras, 432, 1947

\bibitem[{Muñoz-Cuartas {et~al.}(2011)Muñoz-Cuartas, Macciò, Gottlöber, \&
  Dutton}]{Munoz-Cuartas_2011}
Muñoz-Cuartas, J.~C., Macciò, A.~V., Gottlöber, S., \& Dutton, A.~A. 2011,
  \mnras, 411, 584

\bibitem[{{Peter} {et~al.}(2013){Peter}, {Rocha}, {Bullock}, \&
  {Kaplinghat}}]{Peter_2013}
{Peter}, A. H.~G., {Rocha}, M., {Bullock}, J.~S., \& {Kaplinghat}, M. 2013,
  \mnras, 430, 105

\bibitem[{Power {et~al.}(2003)Power, Navarro, Jenkins, Frenk, White, Springel,
  Stadel, \& Quinn}]{Power_2003}
Power, C., Navarro, J.~F., Jenkins, A., {et~al.} 2003, \mnras, 338, 14

\bibitem[{Prada {et~al.}(2019)Prada, Forero-Romero, Grand, Pakmor, \&
  Springel}]{Prada_2019}
Prada, J., Forero-Romero, J.~E., Grand, R. J.~J., Pakmor, R., \& Springel, V.
  2019, \mnras, 490, 4877

\bibitem[{Pulsoni {et~al.}(2020)Pulsoni, Gerhard, Arnaboldi, Pillepich, Nelson,
  Hernquist, \& Springel}]{Pulsoni_2020}
Pulsoni, C., Gerhard, O., Arnaboldi, M., {et~al.} 2020, \aap, 641, A60

\bibitem[{Read \& Gilmore(2005)}]{Read_2005}
Read, J.~I. \& Gilmore, G. 2005, \mnras, 356, 107

\bibitem[{Robertson {et~al.}(2019)Robertson, Harvey, Massey, Eke, McCarthy,
  Jauzac, Li, \& Schaye}]{Robertson_2019}
Robertson, A., Harvey, D., Massey, R., {et~al.} 2019, \mnras, 488, 3646–3662

\bibitem[{Robertson {et~al.}(2022)Robertson, Huff, \&
  Markovic}]{Robertson_2022}
Robertson, A., Huff, E., \& Markovic, K. 2022, arXiv e-prints
  [\eprint[arXiv]{2210.13474}]

\bibitem[{{Sameie} {et~al.}(2018){Sameie}, {Creasey}, {Yu}, {Sales},
  {Vogelsberger}, \& {Zavala}}]{Sameie_2018}
{Sameie}, O., {Creasey}, P., {Yu}, H.-B., {et~al.} 2018, \mnras, 479, 359

\bibitem[{Schneider {et~al.}(2012)Schneider, Frenk, \& Cole}]{Schneider_2012}
Schneider, M.~D., Frenk, C.~S., \& Cole, S. 2012, \jcap, 2012, 030

\bibitem[{Shen {et~al.}(2022)Shen, Brinckmann, Rapetti, Vogelsberger, Mantz,
  Zavala, \& Allen}]{Shen_2022}
Shen, X., Brinckmann, T., Rapetti, D., {et~al.} 2022, \mnras, 516, 1302

\bibitem[{{Springel} {et~al.}(2004){Springel}, {White}, \&
  {Hernquist}}]{Springel_2004}
{Springel}, V., {White}, S.~D.~M., \& {Hernquist}, L. 2004, in Dark Matter in
  Galaxies, ed. S.~{Ryder}, D.~{Pisano}, M.~{Walker}, \& K.~{Freeman}, Vol.
  220, 421

\bibitem[{Springel {et~al.}(2001)Springel, White, Tormen, \&
  Kauffmann}]{Springel_2001}
Springel, V., White, S. D.~M., Tormen, G., \& Kauffmann, G. 2001, \mnras, 328,
  726

\bibitem[{Stadel {et~al.}(2009)Stadel, Potter, Moore, Diemand, Madau, Zemp,
  Kuhlen, \& Quilis}]{Stadel_2009}
Stadel, J., Potter, D., Moore, B., {et~al.} 2009, \mnras, 398, L21

\bibitem[{{Tenneti} {et~al.}(2014){Tenneti}, {Mandelbaum}, {Di Matteo}, {Feng},
  \& {Khandai}}]{Tenneti_2014}
{Tenneti}, A., {Mandelbaum}, R., {Di Matteo}, T., {Feng}, Y., \& {Khandai}, N.
  2014, \mnras, 441, 470

\bibitem[{Tissera {et~al.}(2010)Tissera, White, Pedrosa, \&
  Scannapieco}]{Tissera_2010}
Tissera, P.~B., White, S. D.~M., Pedrosa, S., \& Scannapieco, C. 2010, \mnras,
  no

\bibitem[{Valenzuela {et~al.}(in prep.)Valenzuela, Remus, \&
  Dolag}]{Valenzuela_2022}
Valenzuela, L.~M., Remus, R.-S., \& Dolag, K. in prep., arXiv e-prints

\bibitem[{Vargya {et~al.}(2022)Vargya, Sanderson, Sameie, Boylan-Kolchin,
  Hopkins, Wetzel, \& Graus}]{Vargya_2021}
Vargya, D., Sanderson, R., Sameie, O., {et~al.} 2022, \mnras

\bibitem[{{Velliscig} {et~al.}(2015){Velliscig}, {Cacciato}, {Schaye}, {Crain},
  {Bower}, {van Daalen}, {Dalla Vecchia}, {Frenk}, {Furlong}, {McCarthy},
  {Schaller}, \& {Theuns}}]{Velliscig_2015}
{Velliscig}, M., {Cacciato}, M., {Schaye}, J., {et~al.} 2015, \mnras, 453, 721

\bibitem[{{Warnick} {et~al.}(2008){Warnick}, {Knebe}, \&
  {Power}}]{Warnick_2008}
{Warnick}, K., {Knebe}, A., \& {Power}, C. 2008, \mnras, 385, 1859

\bibitem[{Zemp {et~al.}(2011)Zemp, Gnedin, Gnedin, \& Kravtsov}]{Zemp_2011}
Zemp, M., Gnedin, O.~Y., Gnedin, N.~Y., \& Kravtsov, A.~V. 2011, \apjs, 197, 30

\bibitem[{Zemp {et~al.}(2012)Zemp, Gnedin, Gnedin, \& Kravtsov}]{Zemp_2012}
Zemp, M., Gnedin, O.~Y., Gnedin, N.~Y., \& Kravtsov, A.~V. 2012, \apj, 748, 54

\end{thebibliography}
